\documentclass[twocolumn,showpacs,preprintnumbers,amsmath,amssymb]{revtex4}

\usepackage{graphicx}
\usepackage{dcolumn}
\usepackage{bm}
\usepackage{colordvi}
\usepackage{epsfig}
\usepackage{graphicx}
\usepackage{color}

\newcommand{\kap}{\boldsymbol{\kappa}}
\newcommand{\rb}{{\bf r}}
\newcommand{\bp}{\boldsymbol{\partial}}
\newcommand{\qb}{{\bf q}}
\newcommand{\kb}{{\bf k}}
\newcommand{\pb}{{\bf p}}
\newcommand{\sig}{\boldsymbol{\sigma}}
\newcommand{\ep}{\boldsymbol{\epsilon}}
\newcommand{\F}{\boldsymbol{F}}
\newcommand{\B}{\boldsymbol{B}}
\newcommand{\C}{\boldsymbol{C}}

\begin{document}

\title{A First-Principles Constitutive Equation for Suspension Rheology}

\author{J.~M.~Brader$^1$, M.~E.~Cates$^2$ and M.~Fuchs$^1$}
\affiliation{
$^1$Fachbereich Physik, Universit\"at Konstanz, D-78457 Konstanz, Germany
\\
$^2$SUPA, School of Physics, The University of Edinburgh,
Mayfield Road, Edinburgh EH9 3JZ, UK
}


\begin{abstract}
Using mode-coupling theory, we derive a constitutive equation for the nonlinear rheology of dense 
colloidal suspensions under arbitrary time-dependent homogeneous flow. 
Generalizing previous results for simple shear, this allows the full tensorial structure of the 
theory to be identified. Macroscopic deformation measures, such as the Cauchy-Green tensors, 
thereby emerge. So does a direct relation between the stress and the distorted microstructure, 
illuminating the interplay of slow structural relaxation and arbitrary imposed flow. We present 
flow curves for steady planar and uniaxial elongation and compare these to simple shear. 
The resulting non-linear Trouton ratios point to a tensorially nontrivial dynamic yield 
condition for colloidal glasses.
\end{abstract}

\maketitle 
The imposition of flow can drive a fluid far from
equilibrium. Due to the occurance of long relaxation times, this effect
is ubiquitous in 
complex fluids (colloids, polymers, etc.), whose rheology is of significant technological 
interest, and also represents an important challenge in
nonequilibrium statistical physics. Continuum approaches have provided 
important insights, using symmetry and other principles to construct or constrain
phenomenological constitutive relations. While the
constitutive equations of Newtonian fluids and Hookian solids are
derivable from fundamental starting points (the theory of
linear response based on Onsager's regression hypothesis), 
there has been less progress with their nonlinear generalizations for viscoelastic
fluids, plastic solids and other strongly deforming soft materials. 
\par
A central aim of theoretical rheology is thus to derive from the
underlying microscopic interactions the constitutive equations
that relate the stress tensor to the macroscopic deformation history of a material.
For entangled polymer melts, the constitutive
equation of Doi and Edwards \cite{doi_edwards} has enjoyed considerable success. 
An analogously general microscopic constitutive equation for colloidal dispersions remains 
conspicuously lacking \cite{brady}. 
Even the simplest hard-sphere
colloids in concentrated suspension exhibit a broad range of
viscoelastic behavior; alongside to
flow-thinning \cite{crassous} and thickening
\cite{wagner}, slow structural relaxation leads to glasses showing solid-like response, 
strain-hardening/softening, and plastic flow \cite{pham}. 
\par
But, while the linear viscoelastic spectra of colloidal suspensions are fairly well understood 
\cite{bergenholtz_naegele}, only recently has progress been made in nonlinear flow predictions 
for simple shear \cite{fuchs_cates_PRL,joeprl}.
Shear represents a relatively weak flow in which material lines grow 
linearly with time, while in elongational flows such growth is
exponential, creating much more severe deformations 
of material elements. Thus a description capable of handling arbitrary deformation histories 
is highly desirable.
In the continuum approaches, invariance arguments strongly restrict the deformation measures 
that can appear. A good microscopic theory (e.g., \cite{doi_edwards}) should implicitly 
respect such invariances so that a tensorially admissible constitutive structure emerges 
from a first-principles starting point.
\par
In this Letter we develop such a theory for dense suspensions of spherical
colloidal particles under imposed time-dependent flow. 
Our approach, which considerably generalizes \cite{fuchs_cates_PRL,joeprl}, is nonlinear in the 
velocity gradient tensor and, in the absence of flow, 
features a transition
to a glassy solid as a function of the thermodynamic control
parameters, allowing the delicate interaction between glass
formation and (time-dependent) external deformation
to be investigated.  
Our treatment assumes incompressible homogeneous flow,
neglecting fluctuations of the solvent velocity field and thus
hydrodynamic interactions \cite{fuchs_cates_PRL}. 
These assumptions may break down at high flow rates and/or densities 
where lubrication forces drive cluster formation and shear thickening 
\cite{wagner}. 
The nonlinear theory is formulated by application of
mode-coupling type approximations to exact generalized Green-Kubo
relations. 
We thus consider it to be of `dynamical mean field type'.
\par
We consider a system of $N$ spherical Brownian particles dispersed in a solvent
with a specified velocity profile ${\bf v}(\rb,t)=\kap(t)\cdot\rb$.
The time-dependent velocity gradient tensor $\kap(t)$ is assumed spatially constant,
thus excluding the inhomogeneous flows which occur in shear-banded and shear-localized
states.
Incompressibility implies that ${\rm Tr}\,\kap(t)=0$.
The distribution function evolves according to a Smoluchowski equation 
$\partial_t \Psi(t) = -\sum_i \boldsymbol{\partial}_i\cdot {\bf j}_i$ (where ${\bf j}_i$ is 
the probability current of particle $i$) \protect\cite{dhont,doi_edwards}:
\begin{eqnarray}
&&\partial_t \Psi(t) = \Omega(t) \Psi(t)\notag\\
\Omega(t) =&& \!\!\!\!\!\sum_i \bp_i\cdot[\bp_i - {\bf F}_i - \kap(t)\cdot\rb_i].
\label{smol}
\end{eqnarray}
\vspace*{-0.3cm}\\
${\bf F}_i$ is the force acting on particle $i$ due to the other particles.
The thermal energy $k_BT$ and infinite-dilution diffusion coefficient $D_0$ are 
set equal to unity.
Equilibrium and non-equilibrium solutions to (\ref{smol}) are
distinguished by the existence of a finite probability current. 
Its existence rules out the possibility of a distribution function 
of Gibbs-Boltzmann form.
\par
When constructing constitutive equations based on a given microscopic dynamics 
it is important to first identify the relevant affine deformation measures.  
We achieve this by considering the translational invariance 
of the two-time correlation functions. 
The first step is to show that 
a translationally invariant initial distribution
function can lead to a {\em translationally invariant}, but anisotropic,
distribution function $\Psi(t)$, despite the fact
that the Smoluchowski operator is itself not translationally
invariant. 
This follows from the formal solution 
\begin{eqnarray}
\Psi(\Gamma,t)=e_+^{\int_0^t ds\, \Omega(\Gamma,s)}\Psi_e(\Gamma),
\label{formal_solution}
\end{eqnarray}
where we assume an equilibrium Gibbs-Boltzmann distribution $\Psi_e(\Gamma)$ at 
$t=0$, $\Gamma\equiv \{ {\bf r}_1,\cdots, {\bf r}_N \}$ and $e_+$ 
is a time-ordered exponential \cite{sup_mat}.
We now consider shifting the particle coordinates by a constant vector
${\bf r}'_i = {\bf r}_i + {\bf a}$. This leads to
$\Omega(\Gamma',t)=\Omega(\Gamma,t) - \sum_i \bp_i\cdot\kap(t)\cdot {\bf a}
\equiv\Omega(\Gamma,t) + A(t)$ and hence the shifted distribution function
\begin{eqnarray}
\Psi(\Gamma',t)=e_+^{\int_0^t ds\, (\Omega(\Gamma,s)+A(s))}\Psi_e(\Gamma),
\label{shifted}
\end{eqnarray}
where $\Psi_e(\Gamma')=\Psi_e(\Gamma)$.
The time-ordered exponential can be rewritten using an operator identity
\begin{eqnarray}
\Psi(\Gamma',t)=\exp_+\!\left(
\int_0^t \!\!\!ds\, e_+^{\int_s^t \!ds'A(s')}\Omega(\Gamma,s)
e_-^{\!-\!\int_s^t \!ds'A(s')}
\right)&& \notag\\
&&\hspace*{-4cm}
\times e_+^{\int_0^t \!ds\, A(s)}\Psi_e(\Gamma).
\label{identity}
\end{eqnarray}
Using $A(t)\Psi_e(\Gamma)=0$ and applying the commutator $[\Omega(\Gamma,s),A(s')]
= -{\bf a}\cdot\kap^T(s)\cdot\kap^T(s')\cdot\sum_i \bp_i$ yields
\begin{eqnarray}
\Psi(\Gamma',t)=\Psi(\Gamma,t),
\label{trans_inv}
\end{eqnarray}
as desired.  
We now use this result to study the invariance properties of the
correlation functions.
The correlation of two wavevector-dependent fluctuations is given by
$C_{f_{\qb}g^{}_{\kb}}(t,t')
\equiv\langle\delta f_{\qb}(t)\delta g^{}_{\kb}(t')\rangle_{\kap(t')}$, where
$\langle \,\cdot\, \rangle_{\kap(t')}$ indicates an average over the
distribution (\ref{formal_solution}). Translating the particles 
by a constant vector ${\bf a}$ and using (\ref{trans_inv}) yields
\begin{equation}
C_{f_{\qb}g^{}_{\kb}}(t,t') = e^{-i (\,\bar{\qb}(t,t')-\kb\,)\cdot {\bf a}}
\,C_{f_{\qb}g^{}_{\kb}}(t,t').
\label{invariance}
\end{equation}
Translational invariance requires that the correlation function is unaffected 
by the shift with ${\bf a}$. 
This leads to the requirement 
that a fluctuation at wavevector $\kb=\bar{\qb}(t,t')$ at time $t'$ is correlated
with a fluctuation with wavevector $\qb$ at time $t$ 
as a result of the affine solvent flow, where
\vspace*{-0.2cm}
\begin{eqnarray}
\bar{\qb}(t,t')=\qb\cdot e_-^{-\int_{t'}^{t}\!ds\,\kap(s)}.
\label{advection1}
\end{eqnarray}
\vspace*{-0.5cm}\\
For shear (\ref{advection1}) recovers the familiar shear-advection \cite{joeprl}, while
for an extensional flow (\ref{advection1}) describes the exponential
deformation of material lines.
The tensorial exponential function in (\ref{advection1}) may appear unfamiliar,
but is simply the inverse of the deformation gradient tensor
$\F(t,t')\equiv \partial\rb(t)/\partial\rb(t')$, a standard
quantity in elasticity theory used to connect initial $(t')$ and final $(t)$
coordinates following homogeneous deformation, $\rb(t)=\F(t,t')\cdot\rb(t')$.
The deformation gradient is related to the velocity gradient tensor via
$\partial_t\F(t,t')=\kap(t)\F(t,t')$.
We can thus define the forward and reverse-advected wavevectors
$\bar{\qb}(t,t')$ and $\qb(t,t')$ using the deformation gradient
\begin{eqnarray}
\bar{\qb}(t,t')=\qb\cdot\F^{-1}(t,t'),
\;\;\;\;\;\;\qb(t,t')=\qb\cdot\F(t,t').
\label{advection}
\end{eqnarray}
\par
The magnitudes of the advected wavevectors are thus related 
to the left and right Cauchy-Green tensors, given by $\B(t,t')=\F(t,t')\F^{T}(t,t')$ and  
$\C^{-1}(t,t')=\F^{-1}(t,t')(\F^{-1}(t,t'))^{T}$, respectively \cite{truesdell_noll}:
\begin{eqnarray}
\bar{q}^2(t,t')\!=\!\qb\!\cdot\!\B(t,t')\!\cdot\!\qb\;\;,\;\;
q^2(t,t')\!=\!\qb\!\cdot\!\C^{-1}(t,t')\!\cdot\!\qb\,.
\label{cauchy_green}
\end{eqnarray}
By considering the translational invariance of solutions to (\ref{smol}) we have
thus identified the appropriate deformation measures for describing 
the affine deformation in our Brownian system. 
This enables us to formulate a theory obeying the principle of 
material objectivity; see below and \cite{sup_mat}.
Non-affine particle 
rearrangements arising from excluded volume constraints at finite colloid 
density are approximately treated by the mode-coupling approximations to be 
introduced below.
\par
Integration over the flow history yields an alternative solution of
(\ref{smol}) which is formally equivalent to (\ref{formal_solution}) but
more suitable for approximation.
This integration-through-transients approach yields \cite{fuchs_cates_PRL,joeprl}
\begin{equation}
\Psi(t) = \Psi_e +
\int_{-\infty}^{t}\!\!\! dt'\; \Psi_e\;{\rm Tr}\{\kap(t')\hat{\boldsymbol{\sigma}}\}\;
e_-^{\int_{t'}^t ds\,\Omega^{\dagger}(s)},
\label{distribution}
\end{equation}
where $\hat\sigma_{\alpha\beta}\equiv-\sum_i\,  F_i^{\alpha}r_i^{\beta}$
is the potential part of the microscopic stress tensor, and the adjoint
Smoluchowski operator is $\Omega^{\dagger}(t) = \sum_i
(\bp_i + {\bf F}_i + \rb_i\cdot\kap^T(t))\cdot\bp_i$. 
We take the system to be in quiescent equilibrium in the infinite
past and thus neglect possible non-ergodicity in the initial state
and related ageing phenomena. The validity of this assumption 
will be dependent upon both the specific system and flow
history under consideration \cite{joeprl}. 
By using (\ref{distribution}) to calculate the average of 
$\rho^*_k\rho^{}_k/N$ and, in the spirit of mode-coupling theory, 
approximating by projecting 
onto density fluctuations \protect\cite{goetze_sjoegren,fuchs_cates_PRL} 
we obtain an expression for the distorted structure factor 
$S_{\kb}(t;\kap)=\langle \rho^*_k\rho^{}_k \rangle_{\kap(t)}$
\begin{eqnarray}
&& \!\!\!S_{\kb}(t;\kap) = S_{k}\, -
\int_{-\infty}^t \!\!\!\!\!dt'\, \frac{\partial
S_{k(t,t')}}{\partial t'}\,\Phi^2_{\kb(t,t')}(t,t')
\label{distorted_structure}
\\
\!\!\!\!\!+&&\!\!\!\!\!\!\!
\int_{-\infty}^t \!\!\!\!\!\!dt'\,
\frac{\partial S_{k}}{\partial n}
\!\!\int \!\!\frac{\!d\qb}{16\pi^3}
\frac{\partial S_{q(t,t')}}{\partial t'}
\frac{S_0}{S_q^2}\left( \!S_q \!+ n\frac{\partial S_q}{\partial n}\! \right)
\Phi^2_{\!\qb(t,t')}(t,t'),
\notag
\end{eqnarray}
where $n$ is the number density and $S_k$ the equilibrium structure factor 
used to proxy the colloidal interactions \cite{sup_mat}. 
Eq.(\ref{distorted_structure}) describes the flow induced microstructural 
distortion which becomes appreciable when the flow interferes with slow 
cooperative structural relaxation \cite{szamel}. 
Flow enters the description via the advected wavevector,  
both explicitly through $S_{k(t,t')}$ and 
implicitly through its effect on the transient correlator.
The second term in (\ref{distorted_structure}) is anisotropic, 
whereas the third term is purely isotropic.  
\par
Eq.(\ref{distribution}) also can be employed to directly approximate the 
stress tensor. 
Projection operator approximation of the resulting generalized Green-Kubo 
relation provides an explicit approximation for $\sig(t)$. This we find to be related directly 
to $S_{\kb}(t;\kap)$ by \cite{stress_footnote}
\begin{eqnarray}
\sig(t)=-\Pi\,{\bf 1} -\int \!\!\frac{d\kb\,\rho}{16\pi^3}
\,\frac{\kb\kb}{k}
\,c'_k
\,\delta S_{\kb}(t;\kap),
\label{fl_equation}
\end{eqnarray}
where $c_k$ is the direct correlation function, $c'_k = dc_k/dk$,  
$\delta S_{\kb}(t;\kap) = S_{\kb}(t;\kap) - S_k$ and $\Pi$ is the equilibrium 
osmotic pressure.
(Note that (\ref{fl_equation}) also provides the pressure under flow).
For shear flow $\sigma_{xy}(t)$ from (\ref{fl_equation}) coincides with a result of 
Fredrickson and Larson \cite{fredrickson_larson} for sheared copolymers, reflecting the Gaussian 
statistics underlying both their approach and our own.
Stresses are thus connected to microstructural distortions, which build up over time
via the affine stretching of density fluctuations competing with structural rearrangements 
encoded in $\Phi_{\bf k}(t,t')$.
\par
In order to close our theory we require the transient correlator 
$\Phi_{\bf k}(t,t')\!\!=\!\!\langle\,\rho^*_{\bf k}
\exp_{-}(\int_{t'}^t ds\,\Omega^{\dagger}(s))
\rho_{\bar{\kb}(t,t')} \rangle/N\!S_{k}$ 
describing the decay under flow of thermal density fluctuations, 
where $\langle \cdot \rangle$ denotes an equilibrium average.
The appearance of the advected wavevector ensures that trivial decorrelation 
due to affine motion alone is removed.
Time-dependent projection operator manipulations 
yield exact results for the equation of motion of the transient density correlator 
containing a generalized friction kernel which is amenable to mode 
coupling approximations \cite{fuchs_cates_PRL,joeprl}.
Applying mode coupling type approximations to these formal results
yields the equation of motion
\begin{eqnarray}
\hspace*{0.cm}
\frac{\partial}{\partial t}\Phi_{\bf q}(t,t_0)
&+& \Gamma_{\bf q}(t,t_0)\bigg(
\Phi_{\qb}(t,t_0)
\\
&+&
\int_{t_0}^t dt' m_{\qb}(t,t',t_0) \frac{\partial}{\partial t'} \Phi_{\qb}(t',t_0)
\bigg) =0
\notag
\label{eom}
\end{eqnarray}
where
$\Gamma_{\bf q}(t,t_0)=\bar{q}^2(t,t_0)/S_{\bar{q}(t,t_0)}$. 
The friction kernel $m_{\bf q}(t,s,t_0)$ is the autocorrelation 
function of fluctuating stresses, which in mode coupling are connected to 
structural relaxation as described by the density correlator. In the 
present approximation it is given by
\begin{eqnarray}
\label{approxmemory}
&& \hspace*{-1cm}
m_{\qb}(t,t'\!,t_0) \!\!= \!\!
\frac{\rho}{16\pi^3} \!\!\int \!\! d\kb
\frac{S_{\bar{q}(t,t_0)} S_{\bar{k}(t',t_0)} S_{\bar{p}(t',t_0)} }
{\bar{q}^2(t',t_0) \bar{q}^2(t,t_0)}\\
&\times&
V_{\qb\kb\pb}(t',t_0)\,V_{\qb\kb\pb}(t,t_0)\Phi_{\bar{\kb}(t',t_0)}(t,t')
\Phi_{\bar{\pb}(t',t_0)}(t,t'),
\notag
\end{eqnarray}
\begin{eqnarray}
\!\!\!\!\!V_{\qb\kb\pb}(t,t_0) \!=\! \bar\qb(t,t_0)\cdot(
\bar\kb(t,t_0) c_{\bar{k}(t,t_0)} \!+
\bar\pb(t,t_0) c_{\bar{p}(t,t_0)}),
\label{vertex}
\end{eqnarray}
where $\pb=\qb-\kb$.
Equations (\ref{distorted_structure})-(\ref{vertex}) form a closed constitutive 
theory for the microstructure and stress response  
of Brownian particles under external flow, requiring only 
$S_k$ and $\kap(t)$ as input \cite{sup_mat}. 
\par
We next consider our theory from the standpoint of 
continuum `rational mechanics' approaches \cite{truesdell_noll}, showing that 
(in common with \cite{doi_edwards}) it complies with their invariance 
requirements but avoids their oversimplifying assumptions. Such approaches often express 
$\sig(t)$ as a functional of a suitable deformation measure, weighted
by a fading memory; an example is
the integral form of the upper-convected Maxwell equation (Lodge equation) with the 
memory taken as a known sum of decaying exponentials \cite{larson}.
In the present treatment, the memory is instead given by $\Phi^2_{\kb}(t,t')$
which for general flows is both anisotropic and a function of two times
rather than a simple time difference. 
Both aspects are physically reasonable extensions of the assumption 
of a fixed isotropic decay.
In the absence of flow, $\phi_{\kb}(t)$ exhibits non-exponential
decay and, in the glass, arrests to a finite plateau value
leading to solid-like response \cite{goetze_sjoegren}.
\par
An important symmetry consideration is the
{\em principle of material objectivity} (or frame indifference) 
\cite{larson,truesdell_noll,material_objectivity}.
This asserts that the relationship between
$\sig$ and $\kap$ should be invariant with
respect to time-dependent rotation of either the material sample or the
observer.
Although this is only an approximate symmetry, based on the
neglect of inertial effects at the microscopic level \cite{material_objectivity},
many soft materials display this invariance to a good level of approximation.
The overdamped Smoluchowski dynamics (\ref{smol}) underlying our treatment
excludes inertial effects from the outset, so that
our set of equations (\ref{distorted_structure})-(\ref{vertex}) are material objective, so long as this is preserved by our approximations. This can be explicitly
confirmed by using (\ref{advection}) and (\ref{cauchy_green}) to eliminate
the advected wavevectors from (\ref{distorted_structure})-(\ref{vertex}) in favour of
the deformation tensors \cite{sup_mat}.
\par
Next we address the comparison between elongational (planar or uniaxial) flow, and shear. Starting with the limit of small strain, we obtain the linear response result
\begin{eqnarray}
\sig^{l}(t)\!=\!\!\int_{-\infty}^{t} \!\!\!\!\!\!dt'\!\int\!\!\!
\frac{d{\bf k}}{16\pi^3} \{(\kb\!\cdot\!\bar{\kap}(t')\!\cdot\!\kb)\kb\kb\}\!
\left(\!\frac{S'_k\Phi_k(t\!-\!t')}{k S_k}\!\right)^{\!2}
\label{linear}
\end{eqnarray}
where linearity enables us to introduce the symmetrized 
tensor $\bar{\kap}(t)=(\kap(t)+\kap^T(t))/2$,
and $\Phi_k(t)$ is the quiescent correlator.
As all anisotropy in (\ref{linear}) is contained within the factor $\{\cdot\}$,
the angular integrals may be easily evaluated for a given $\kap(t)$.
It follows that the planar-extensional and shear viscosities 
are related by $\eta_{\rm e}/\eta_{\rm s}=4$, 
in compliance with one of {\em Trouton's rules}.
In the glass phase the transient correlator does not relax to zero. 
Partial integration of (\ref{linear}), followed by taking the small-strain
limit, leads to $\sig(t)=2G\ep(t)$, where $\ep(t)$ is the infinitesimal strain tensor 
(${\rm Tr}\,\ep(t)=0$), 
$G =\rho^2\int_0^{\infty}\!\!d\kb\, k^4(c'_k\phi_k(\infty))^2/(60\pi^2)$ 
is a known mode-coupling approximation for the linear shear modulus 
\cite{bergenholtz_naegele,goetze_sjoegren}.
\begin{figure}
\hspace*{-0.5cm}
\includegraphics[width=6.65cm,angle=0]{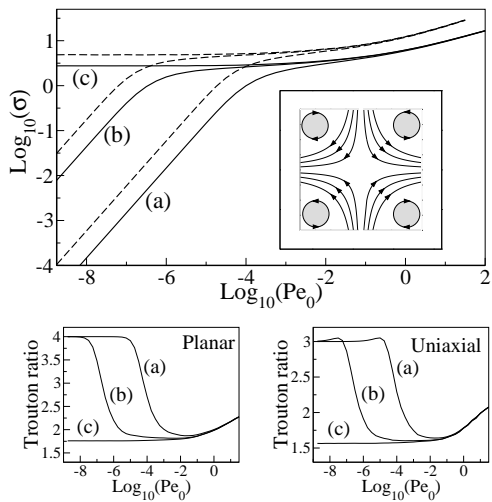}
\caption{Steady state stress $\sigma\!=\!\sigma_{xy}$ under shear
(full lines, $\kappa_{ij}\!=\!\dot\gamma\,\delta_{xi}\delta_{yj}$)
and stress difference $\sigma\!=\!\sigma_{xx}\!-\!\sigma_{yy}$ under planar 
elongation (broken lines, $\kappa_{ij}\!=\!\dot\gamma\,(\delta_{xi}\delta_{xj}
-\delta_{yi}\delta_{yj})$)
as a function of ${\rm Pe}_{0}=\dot\gamma d^2/D_0$, 
where $d$ is the sphere diameter.
Each curve is labelled according to the distance from the glass
transition $\phi - \phi_{c}$, (a) $-10^{-4}$, (b) $\!-10^{-3}$
and (c) $\!10^{-4}$.
The inset shows a possible realization of planar elongational flow.
The lower left panel shows the Trouton ratio for such flow 
while the lower right panel shows the ratio for uniaxial elongation 
($\kappa_{ij}\!=\!\dot\gamma\,(\delta_{xi}\delta_{xj}
-\delta_{yi}\delta_{yj}/2 -\delta_{zi}\delta_{zj}/2)$). 
 (Legend (a-c) defined above.)
}
\label{figure1}
\end{figure}
\par
Taking this comparison beyond linear response requires numerical solution of  
(\ref{distorted_structure})-(\ref{vertex}).
For simplicity we perform calculations for
a one-component system of hard spheres \cite{iso_footnote}.
In Fig.~\ref{figure1} we show
$\sigma_{xy}$ under shear and $(\sigma_{xx}\!-\sigma_{yy})$
under planar elongation as a function of Peclet number, ${\rm Pe}_0=\dot\gamma d^2/D_0$, where
$\dot\gamma$ is the strain rate,
for various packing fractions around the glass transition $\phi_c$. 
A detailed comparison of theoretical shear flow curves with experiment 
can be found in \cite{crassous}; we do not know of similar experimental data on 
elongational flows but hope our work may stimulate future studies.
The lower left panel shows the Trouton ratio
$(\sigma_{xx}\!-\sigma_{yy})/\sigma_{xy}$
as a function of ${\rm Pe}_0$.
Upon entering the nonlinear (shear thinning) regime, this ratio is significantly reduced 
below the linear response value, indicating enhanced strain thinning under elongational 
flow relative to shear.
Moreover, as $\phi_c$ is approached from below, the linear response regime moves to
lower values of ${\rm Pe}_0$ and disappears entirely on crossing the glass
transition. 
The resulting plateau in the flow curve identifies a dynamical yield stress.
The Trouton ratio displays an approximately constant (noninteger) value over the
entire plateau: its low ${\rm Pe}_0$ limit furnishes a nontrivial yield condition for the ratio between normal and shear stresses. 
In the lower right panel of Fig.~\ref{figure1} we show in addition the Trouton ratio for uniaxial elongational flow. 
We again obtain a nontrivial yield stress ratio, distinct from that found 
under planar elongation.  
Our results for uniaxial and planar extension, alongside 
those for shear \cite{fuchs_cates_PRL} thus predict by first-principles theory three 
parameters of a dynamic yield surface for glasses. 
The limiting stress at zero flow rate for a fixed flow geometry identifies 
a point on a dynamic yield manifold. In a system where (as here) isotropic 
pressure is irrelevant, this manifold resides in the space of deviatoric 
(i.e. traceless) stress tensors, and in the principal stress frame (with 
stress eigenvalues $s_{1,2,3}$) can be represented, for instance, as a 
closed curve, $f(s_1-s_3,s_2-s_3) = 0$. Our new results for uniaxial and 
planar extension, alongside those for shear [7], predict from first 
principles three parameters of the dynamic yield manifold for a glass. We 
plan a fuller examination of this manifold in future work.
\par
In conclusion, we have derived from first principles a constitutive equation for dense
colloidal suspensions subjected to an arbitrary time-dependent (but homogeneous) deformation. Appeal to translational invariance identifies appropriate 
deformation measures for an approximate description of the system in terms of density fluctuations.
Within our treatment macroscopic stress is directly connected with structural distortion 
at the microstructural level. 
The inclusion of non-exponential slow relaxation arising from interactions 
leads to a unified description of fluid and glassy states.
For steady shear, planar and uniaxial elongational flows we find 
strong nonlinearities in the Trouton ratio as a function of strain rate. 
For colloidal glasses, we predict nontrivial relations between dynamic yield 
stresses in different flow geometries.
\par
We thank Th.Voigtmann, P.Warren and A.Zippelius for stimulating discussions and
acknowledge the Transregio SFB TR6, EPSRC/GR/S10377 for financial support. 
MEC holds a Royal Society Research Professorship.


\begin{thebibliography}{}
\bibitem{stress_footnote}
Unlike other stress tensor components, for $\Pi$ the agreement between 
the direct approximation to $\sig(t)$ and 
(\ref{fl_equation}) is not exact, but 
a good approximation.
\bibitem{iso_footnote}
In order to approximately solve the equations 
(\ref{distorted_structure})-(\ref{vertex}) we approximate the advection using 
$k^2_{\rm is}(t,t')=\int \!d\Omega\, k^2(t,t')/4\pi$ which enables the angular 
integrals to be performed analytically.
This technical approximation has been successfully applied to
similar problems \cite{faraday_discussions}.
\end{thebibliography}
\end{document}